\documentclass[10pt,journal,compsoc]{IEEEtran}
\ifCLASSINFOpdf
\else
\fi
\usepackage[space,nocompress]{cite}
\usepackage{graphicx,caption,amssymb,amsmath,array}
\usepackage{color,float}
\usepackage{subfigure,epsfig}
\usepackage{ragged2e}
\usepackage{enumitem}
\usepackage{footnote}

\usepackage{algorithm}
\usepackage{algorithmic}
\usepackage{multirow}
\usepackage{float}
\usepackage{url}
\usepackage{enumitem}
\usepackage{array}
\usepackage{booktabs}
\usepackage{bm}
\usepackage{color}

\setlist[itemize]{leftmargin=*}
\setlist[enumerate]{leftmargin=*}
\begin{document}
%
\title{Attention-based Multimodal Feature Representation Model for Micro-video Recommendation}


\author{Mohan Hasama, Jing Li
}

\markboth{Journal of \LaTeX\ Class Files,~Vol.~14, No.~8, August~2015}%
{Shell \MakeLowercase{\textit{et al.}}: Bare Demo of IEEEtran.cls for IEEE Transactions on Magnetics Journals}
%



\IEEEtitleabstractindextext{%
\justifying  
\begin{abstract}
In recommender systems, models mostly use a combination of embedding layers and multilayer feedforward neural networks. The high-dimensional sparse original features are downscaled in the embedding layer and then fed into the fully connected network to obtain prediction results. However, the above methods have a rather obvious problem, that is, the features directly input are treated as independent individuals, and in fact there are internal correlations between features and features, and even different features have different importance in the recommendation. In this regard, this paper adopts a self-attentive mechanism to mine the internal correlations between features as well as their relative importance. In recent years, as a special form of attention mechanism, self-attention mechanism is favored by many researchers. The self-attentive mechanism captures the internal correlation of data or features by learning itself, thus reducing the dependence on external sources. Therefore, this paper adopts a multi-headed self-attentive mechanism to mine the internal correlations between features and thus learn the internal representation of features. At the same time, considering the rich information often hidden between features, the new feature representation obtained by crossover between the two is likely to imply the new description of the user likes the item. However, not all crossover features are meaningful, i.e., there is a problem of limited expression of feature combinations. Therefore, this paper adopts an attention-based approach to learn the external cross-representation of features.
\end{abstract}

\begin{IEEEkeywords}
Micro-video, Recommender Systems, Deep Learning.
\end{IEEEkeywords}}

\maketitle

\IEEEdisplaynontitleabstractindextext

%
\IEEEpeerreviewmaketitle

\section{Introduction}
With the popularization of mobile Internet terminals, the speed of network and the reduction of traffic tariff, people can watch videos through cell phones at any time. As micro-videos are short and rich in content, they can maximize people's demand for watching videos in fragmented time, so micro-videos that combine filming techniques, music, stories and images can meet users' content consumption needs in a more diverse way. Due to the widespread popularity of micro-videos, many micro-video platforms have been born as a result. Micro-video platforms attract users and optimize user experience, micro-video recommendation algorithms have become an important means of competition between platforms, and how to ensure both the accuracy and the real-time nature of the recommendation algorithm has been the focus of research. Along with the rapid popularity of smartphones and mobile Internet, micro-videos, as a new type of user-generated content, have widely appeared on various social platforms, such as Tiktok, kwai, Instagrm, etc.

The existing studies related to content analysis of micro-videos mainly include field Scene estimation~\cite{zhang2016shorter}, popularity prediction of micro-videos low and recommendation of micro-videoss~\cite{liu2019user,shang2016micro}. For example, Zhang et al.~\cite{zhang2016shorter} used text, audio and visual modal features of micro-videos to solve the problem of multimedia scene classification.Wei et al.~\cite{NMCL} used Neural Multimodal Cooperative Learning (NMCL) to solve the problem of micro- video scene classification.Nie et al.~\cite{nie2017enhancing} worked on Jing et al.~\cite{jing2017low} solve the problem of micro-video popularity prediction by proposing a novel low-rank multi-view learning framework. Shang et al.~\cite{shang2016micro} proposed a recommendation system for micro-video big data to achieve recommendation.

In video recommendation, besides applying some basic interaction data (including image data, behavioral data, contextual data, etc.) to the recommendation model, many researches also try to apply the video own content data to the model for video recommendation. As a kind of unstructured data, the rich content and various expression forms of video bring certain challenges to the recommendation system. Currently, video recommendation techniques can be broadly divided into: collaborative filtering-based video recommendation~\cite{huang2016real}, content-based video recommendation~\cite{mei2011contextual,zhu2013videotopic}, and hybrid video recommendation~\cite{ferracani2016item,jiang2020aspect}. The collaborative filtering-based approach uses historical behavior logs and calculates user similarity or video similarity to predict the videos that users may be interested in.

YouTube~\cite{baluja2008video} proposed in 2008 to build the recommendation problem into the network results of user-videos by wandering to find videos of interest, which is essentially an item-based collaborative filtering approach. The collaborative filtering approach is computationally simple and interpretable, but faces the cold-start problem, where it is no longer useful when new videos are added to the library or new users join. An effective way to solve the cold-start problem is to find videos of interest by analyzing video information such as tags, video text, audio, and visuals~\cite{GRCN}. Mei et al.~\cite{mei2007videoreach} combined modal content such as text, audio, and video with attention to make recommendations based on video content and user interaction behaviors, which is limited by considering only video-to-video correlations and ignoring the important information of user preferences. The content-based recommendation system proposed by Deldjoo et al. ~\cite{deldjoo2016content} extracts a series of features from videos such as color, motion, and illumination to analyze video content. Although this approach effectively solves the cold-start problem and enriches video descriptions, improves model prediction. However, it also poses problems of computational difficulties and high video analysis costs for long videos, however, it is still an efficient approach for micro-videos. In recent years, micro-videos have become increasingly popular, and more and more researchers are focusing on micro-video content analysis. Ma et al.~\cite{ma2018lga} proposed an LGA model to input the extracted user-object interaction features, as well as contextual and visual auxiliary features describing the micro-video content, into a neural network to calculate the prediction score. Huang et al.~\cite{huang2017personalized} proposed personalized micro-video recommendations from videos with multiple modal features and modeling user interests from multiple dimensions.

In recommender systems, models mostly use a combination of embedding layers and multilayer feedforward neural networks. The high-dimensional sparse original features are downscaled in the embedding layer and then fed into the fully connected network to obtain prediction results. However, the above methods have a rather obvious problem, that is, the features directly input are treated as independent individuals, and in fact there are internal correlations between features and features, and even different features have different importance in the recommendation. In this regard, this paper adopts a self-attentive mechanism to mine the internal correlations between features as well as their relative importance. In recent years, as a special form of attention mechanism, self-attention mechanism is favored by many researchers. The self-attentive mechanism captures the internal correlation of data or features by learning itself, thus reducing the dependence on external sources. Therefore, this paper adopts a multi-headed self-attentive mechanism to mine the internal correlations between features and thus learn the internal representation of features. At the same time, considering the rich information often hidden between features, the new feature representation obtained by crossover between the two is likely to imply the new description of "the user likes the item". However, not all crossover features are meaningful, i.e., there is a problem of limited expression of feature combinations. Therefore, this paper adopts an attention-based approach to learn the external cross-representation of features. The main contributions of this paper are as follows:
\begin{itemize}
    \item In this paper, we propose an attention-based multimodal feature fusion method to assign different weights to each modality of the video through an attention mechanism. In order to ensure that the learned weights of each modality can reflect the user's preferences, the method incorporates user features into the attention network to achieve personalized fusion of the features of each modality of the video, thus providing more personalized video recommendations. 
    \item To address the main micro-video classification tasks, by establishing similarity loss and difference loss, we explore the similarity between different modalities in micro-videos and the difference of the same modality, to obtain the private domain features and public domain features of different modalities of the video, and fuse them as the global features, and the classification loss is used to guide the classification of micro-videos.
    \item Extensive experiments conducted on a real-world dataset have well-verified that our model significantly outperforms several state-of-the-art baselines.

\end{itemize}

\begin{figure}
	\centering
	  \includegraphics[width=0.5\textwidth]{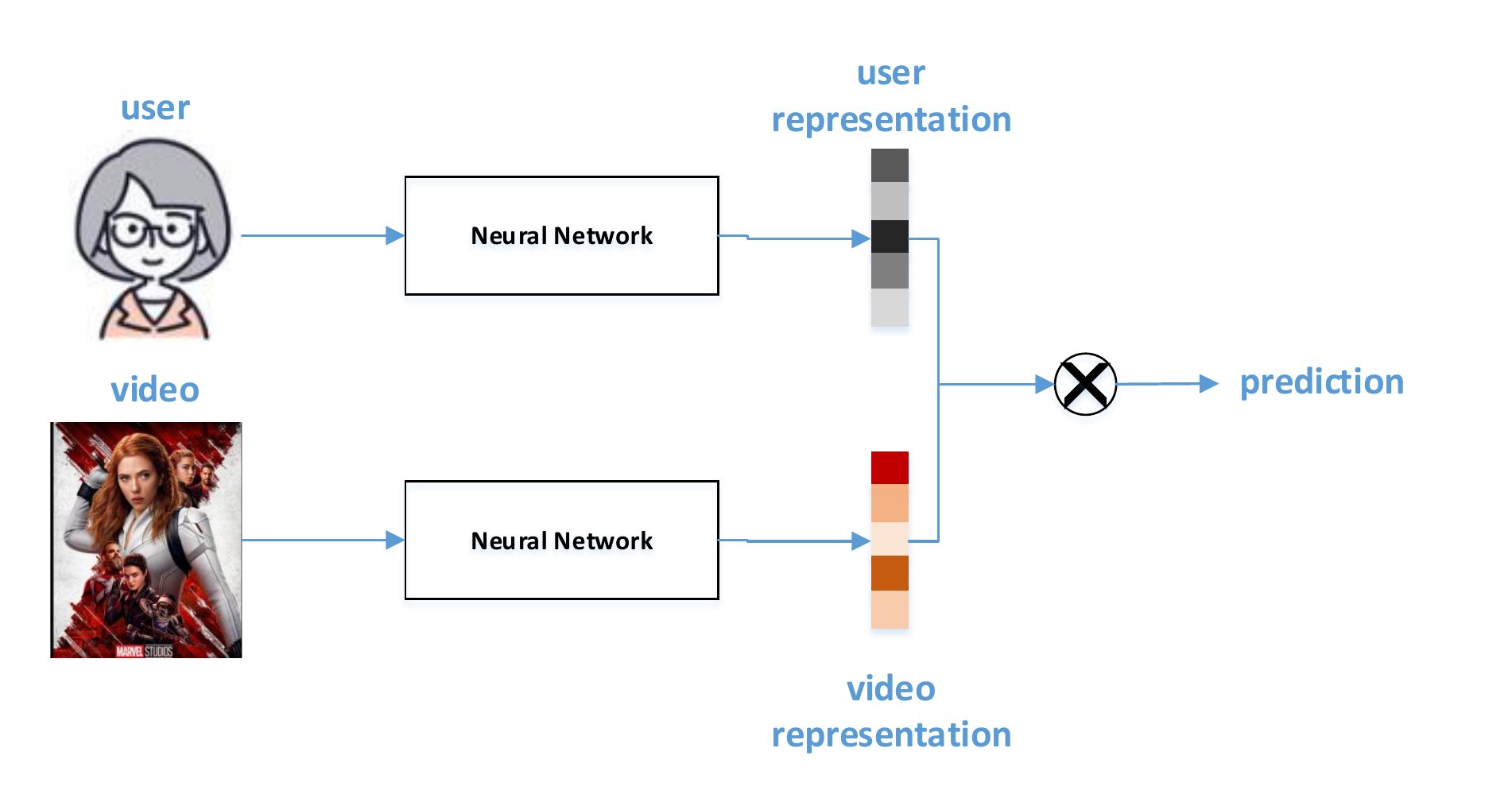}
	  \vspace{-1mm}
	  \caption{Demonstration of representation in recommender system.}
	\label{fig_1}	  
	\vspace{-5mm}
\end{figure}

\section{Related Work}
\subsection{Recommender System}

Traditional recommendation algorithms can be divided into three main categories: Collaborative Filtering (CF)~\cite{nakamura1998collaborative}, Content-Based (CB) ~\cite{pazzani2007content}, and hybrid recommendation algorithms~\cite{balabanovic1997fab}. User-Based Collaborative Filtering algorithm (User-Based CF) was proposed in 1992~\cite{goldberg1992using} and is the earliest algorithm in recommendation systems. The main idea of User-Based CF is that similar user populations have similar preferences. Therefore, the User-Based CF algorithm includes two main steps: first, finding the set of similar users based on their historical behavior, and second, finding the items that the target users like and have not interacted with based on this similarity~\cite{MMGCN}.

MatrixFactorization model (MF)~\cite{koren2009matrix} is one of the Model-Based CF algorithms. The core idea of MF is to link user interest and item features through implicit features and decompose the user-item matrix into the product of two matrices, i.e., user interest matrix and item attribute matrix. The output of this model is a vector of user interests and an item attribute vector with the same dimensionality. The Content-Based (CB) recommendation algorithm, on the other hand, uses the content features or tags of the items themselves to predict the user's interests and thus recommend the content of interest to the user. Since CB algorithms rely on the content features of items, there is no "cold start" problem for new items~\cite{wei2021contrastive}. However, for unstructured data such as video, music, and images, the cost of extracting content features is very expensive. GBDT is a combination of collaborative filtering recommendation and content-based recommendation, each of which has its own strengths to provide more effective recommendations, and the way they are mixed differs in different application scenarios. GBDT generates new features by automatically combining features, and feeds the newly generated features into the LR linear model to obtain predicted results~\cite{he2014practical}.

To explore the cross-correlation of features, Rendle~\cite{rendle2010factorization} proposed Factorization Machines (FM), which can automatically combine features in two, thus mining the implicit information between features and improving the recommendation performance of the model. Recommendation models based on deep learning can be broadly classified into two categories: deep models based on Representationlearning and deep models based on Matchfunctionlearning. Sedhain et al.~\cite{sedhain2015autorec} proposed a self-encoder (auto-encode) based collaborative filtering model. This model combines collaborative filtering and auto-encode by taking each row or column of the scoring matrix as input and learning low-dimensional vector representations of users and items using the Encode and Decode processes. Xue et al.~\cite{xue2017deep} proposed the DMF model by adding a multilayer perceptron (MLP) to the traditional MF model, and learning the representation of users and items through the MLP. Since the input user or item vector is represented by the interacted item id or user id, which is the one-hot encoding form. When the user size is too large (assume 1 million), setting the number of nodes of the first layer perceptron to 100, the number of first layer parameters alone reaches 100 million. The higher the number of parameters of the model, the more likely the model will be overfitted, and therefore the larger the amount of data required to train the model. Kim et al.~\cite{kim2016convolutional} proposed a convolutional matrix factorization model (ConvMF) to obtain a representation of items through a convolutional neural network (CNN). In addition, in multimedia recommendation, Chen et al.~\cite{chen2017attentive} added attention mechanism to the traditional collaborative filtering, and in the paper, two layers of attention are used to learn the feature representations of users and items, the first layer of attention believes that different weights should be assigned to items that users have interacted with historically; the other layer of attention lies in assigning different weights to multimedia features in the same item.

\begin{figure*}
	\centering
	  \includegraphics[width=1\textwidth]{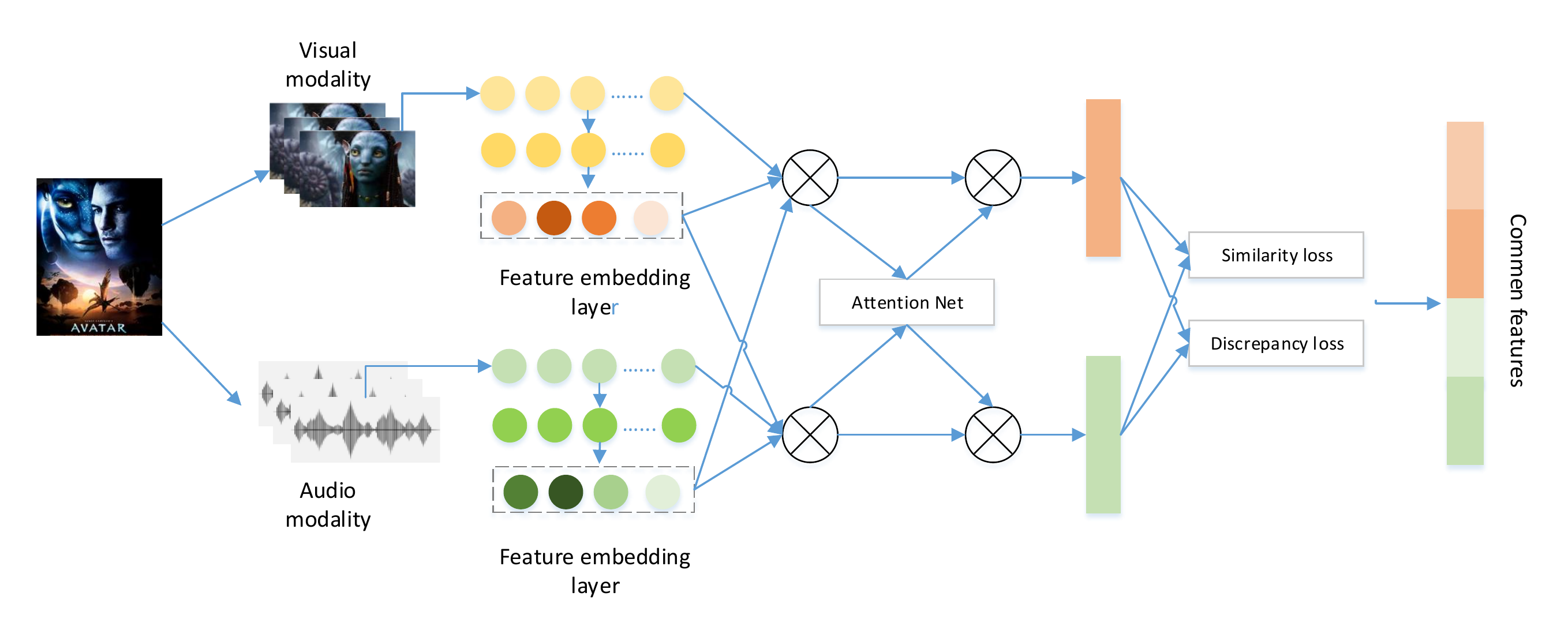}
	  \vspace{-1mm}
	  \caption{Schematic illustration of our proposed model.}
	\label{fig_1}	  
	\vspace{-5mm}
\end{figure*}

\subsection{Feature Extraction}

Early video feature extraction methods use 2D convolutional networks to learn features of each frame of a video, which is inspired by image processing. However, it ignores the temporal correlation between consecutive video frames~\cite{tao2020mgat}. In order to preserve temporal correlation, many existing methods represent the video features as a whole by aggregating the frame features of the video. Long et al.~\cite{long2018attention} proposed Attention ClusterNetwork (ACN), which clusters the local features of the video into the global features of the video by employing attention units. Ma et al.~\cite{ma2019ts} used the attention clusters to achieve feature fusion by setting the features of each moment and the previous moment weighted by weights to realize video classification. In recent years, in order to make full use of the spatio-temporal features of video, 3D convolutional networks have been proposed to learn continuous frame features of video instead of single video frame features. The input parameters of 3D convolutional networks retain the four parameters of video batch size, video channel, video frame width and video frame height, and add the parameter of video depth to record the number of video Tran et al.~\cite{tran2015learning} proposed C3D network to extract the spatio-temporal domain features of continuous frame sequences using 3D convolution and achieved a great breakthrough in video classification accuracy. In recent years, a series of video feature extraction methods introduced based on 3D convolution have been widely used in video classification, tracking, segmentation, etc. For example, Carreira et al.~\cite{carreira2017quo} proposed 3D network to improve the network classification performance by increasing the network width. Hara et al. ~\cite{hara2018can} extended the ResNet originally applied to 2D convolutional network to 3D convolutional network by proposing ResNet3D to solve the problems related to video classification. Feichtenhofer et al.~\cite{feichtenhofer2019slowfast} proposed SlowFastNetwork, in which the whole network was constructed by building two 3D convolutional networks to obtain the global features of the video. However, compared with the traditional 2D convolutional network, the 3D convolutional network requires a larger number of parameters and storage space. To solve this problem, Qiu et al.~\cite{qiu2017learning} constructed a P3D network by combining 3D convolution kernels with 2D convolution in the spatial domain and 1D convolution in the temporal domain. Tran et al.~\cite{tran2018closer} used $R2 + 1D$ networks to decompose 3D convolutional networks into separate spatial and temporal modules. Xie et al.~\cite{xie2017rethinking} used $( 2D + 1D)$ convolutional kernels instead of convolutional kernels in S3D-G networks.
\section{Methodology}
The feature representation model proposed in this paper has following main structures: input layer, embedding layer, MHSA layer, AC layer, and output layer. In the embedding layer, the discrete and sparse features are embedding, and the embedding matrix is used to reduce the dimensionality and learn the original feature representation with more generalization. The MHSA layer and AC layer are two-layer feature representation modules, where the MHSA (Multi-Head-Self-Attention) layer is a feature internal representation module based on Multi-Head Self-Attention, which learns the internal correlation between features through the Multi-Head Self-Attention mechanism, and the multi-head approach The AC (Attention-Crossing) layer is an attention-based feature external cross-representation module, which learns the implicit feature descriptions through the multi-head self-attention mechanism, and learns the importance of feature crossings by means of attention. The problem of limited feature cross-representation is thus solved. In the output layer, the feature internal representation obtained from the MHSA layer and the feature cross representation obtained from the AC layer are weighted and summed, and then the output result is obtained by the sigmoid function. The final output is expressed as:
\begin{equation}
y_b=\delta(W^T_{MSHA}S_{MSHA}+W^T_{AC}P_{AC}+b),
\end{equation}
where $\delta(\cdot)$ represents sigmoid function, $S_{MSHA}$ is the output of $MSHA$,i.e. inner feature representation. $P_{AC}$ is the output of $PA$, i.e. external feature representation.

Considering that our model is always a low-order feature representation, this paper learns higher-order feature information by matching the deep part. The designed structure is shown in Figure 2. The output of the MHSA layer $S_{MSHA}=[s_1, s_2, \dots, s_n]$ in our two-layer model is combined with the output of the AC layer $P_{AC}=[p_1, p_2, \dots, p_n]$, and then used as input to the multilayer perceptron to learn higher order feature expressions. The output of the multilayer perceptron is finally used as the prediction result by the sigmoid function, and the calculation process is shown below:
\begin{equation}
a^{(0)}=concat(S_{MSHA}, P_{AC})=[c_1,c_1, \dots, c_m],
\end{equation}
where $\delta(\cdot)$ is the sigmoid function, $x$is the original input feature, and $w_i$ denotes the weight of the i-th input feature of the linear part. $DNN(a^{(0)})$ represents the result of the depth part of the model, $a^{(0)}$ is the input vector of the depth part, and $a^{(0)}$ is composed of the output of the MHSA layer and the AC layer for stitching. $DNN(a^{(0)})$ can be further decomposed as:
\begin{equation}
DNN(a^{(0)})=FCS(a^l)=W^l a^{(l-1)}+b^l,
\end{equation}
where $FCS(\cdot)$ is the fully connected function representation, $a^l$ denotes the output result of the l-th fully connected layer, $W^l$, $b^l$ are the model trainable parameters that represent the weights and biases of the fully connected layer of the l-th layer.
\begin{figure*}
	\centering
	  \includegraphics[width=1\textwidth]{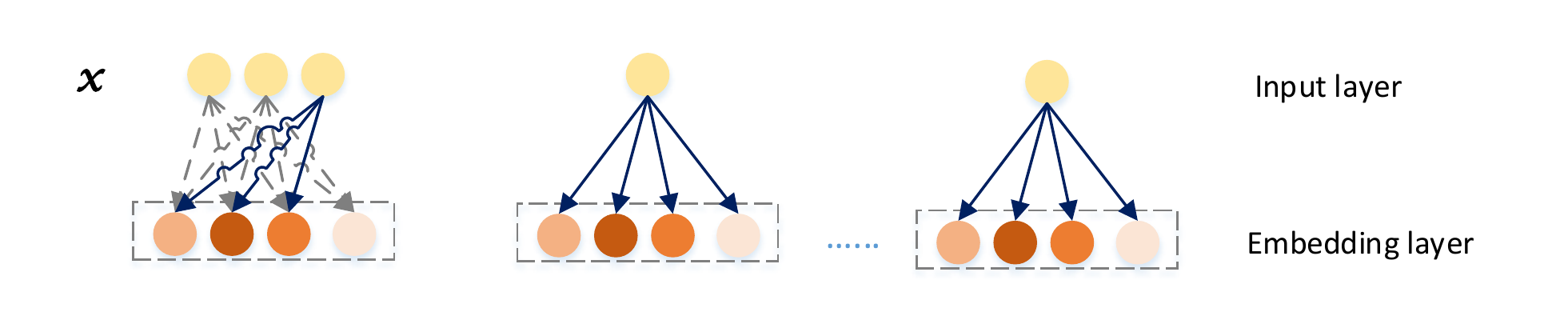}
	  \vspace{-1mm}
	  \caption{Illustration of embedding layer.}
	\label{fig_1}	  
	\vspace{-5mm}
\end{figure*}

\subsection{Feature Extraction}

In order to extract the time-domain information of the visual and audio modalities of the micro-video, this paper uses 3D convolutional networks to obtain the private domain features of the visual and audio modalities and the public domain features of the visual and audio modalities, respectively. For the visual modality, a sequence of 32 consecutive video frames is extracted from the micro-video with the size of $224\times224$; for the audio modality, the micro-video is first divided into 32 video clips at equal intervals from the whole micro-video, and then the audio clips of these 32 video clips are extracted and converted into a spectrogram to represent the change pattern of this audio segment. It should be noted that the spectrogram has only a single channel compared to the video frames. By comparing the accuracy and training complexity of different 3D convolutional networks, we decided to use the I3D network as the feature extraction network, and fine-tuned the output channels of the network, added the average pooling layer, and removed the final fully connected layer for classification. In the overall structure of the network, the whole network is divided into three modules according to the feature extraction perspective: private domain network for visual modal information, public domain network for audio-visual modal information, and private domain network for audio modal information. The parameters of the network model are optimized by reducing the value of the overall loss function to achieve feature extraction and classification of micro-videos. The implementation of the loss function consists of two parts: (i) similarity loss LS, which is used to explore the similarity between different modalities (i.e., public domain features of visual modal information and public domain features of audio modal information); (ii) difference loss LD, which is used to measure the difference within the same modality, i.e., between private and public domain features of visual modal information and between private and public domain features of audio modal information. 

\subsection{Feature Embedding}

In practical applications, the extracted features are often high-dimensional and sparse. After the embedding layer, the high-dimensional sparse feature vectors can be embedded to low-dimensional dense vectors, reducing the dimensional disaster while making the model more robust. The embedding layer is actually a fully connected layer. A high-dimensional sparse feature vector for the input layer is shown below:
\begin{equation}
x_{input}=[x_1, x_2, \dots, x_n],
\end{equation}
where $x_i$ is denoted as the class i feature, and if the class i feature is a discrete (category) feature, then $x_i$ is the one hot coded vector, and if it is a continuous (numerical) feature, then $x_i$ is the normalized eigenvalue. The discrete feature vector is linearly transformed to another feature space by multiplying it with the embedding matrix as follows:
\begin{equation}
x_{emb\_i}=W_i x_i,
\end{equation}
where $x_i$ is the single category one hot encoding vector, $W_i \in R^{N\times D}$ is the embedding matrix, and $d$ is the feature embedding dimension. For multi-valued category features, e.g., action movie , comedy movie, the feature vector is represented as $[1,1,0,0]$ and the embedded features are represented as the average value after the embedding matrix as follows:
\begin{equation}
x_{emb\_i}=\frac{1}{q} W_i x_i,
\end{equation}
where $x_i$ is the multiclass one hot encoding vector, $W_i \in R^{N\times D}$  is the embedding matrix, and $q$ is the number of 1s in the multiclass one hot encoding the number of 1's in the vector. The continuous eigenvalues are converted to equal dimensional eigenvectors by multiplying them with the embedding vectors, as shown in Eq.5, to ensure that each feature The cross-combination between the features can be performed to obtain a more effective feature representation.

\subsection{Feature Representation}

In the recommendation system, the input of original features are independent of each other, ignoring the interaction information of features. The cross-relationship between features is more of an "and" relationship rather than an "add" relationship. And the cross-features of this "and" relationship contain rich information content, which is very important in practical application scenarios. Xiao et al.~\cite{xiao2017attentional} proposed an AFM model to learn the weights of crossover features by introducing attention networks. The feature external cross-representation module designed in this paper draws on this. In this module, the primary task is to crossover the features to the second order, and n embedded feature vectors are multiplied two by two to obtain $n(n-1)^2$ second-order crossover vectors. Since it is an element-level multiplication, the dimensionality of the obtained crossover vectors remains the same, and the definition of the crossover function is shown as follows:
\begin{equation}
\phi_{i,j}(x)=x_i \odot x_j,
\end{equation}
where $\odot$ denotes the vector element-level multiplication,  and $x_i$ denote the embedded feature vectors, and $\phi_{i,j}(x)=x_i $ is the intersection of the feature $x_i$  and  $x_j$ represented by the cross vector. Second, by constructing an attention network  to calculate the importance of each intersection feature weight, the calculation process is shown as follows:
\begin{equation}
M_{i,j}=\frac{M^{'}_{i,j}}{\sum_{(i,j)\in R_x} M^{'}_{i,j}},
\end{equation}

\begin{equation}
M^{'}_{i,j} = h^T ReLU(W\phi_{i,j}(x)+b),
\end{equation}
where $R_x$ denotes all combinations of second-order crossover features.

\subsection{Loss Function}

The twin similarity loss proposed by Chopra et al.~\cite{chopra2005learning}, which was mainly applied in the field of face recognition and achieved good results. Based on this more and more people have devoted to the optimization and use of twin network structure. Zagoruyko and Komodakis~\cite{zagoruyko2015learning} optimized the twin network and applied it in image restoration. Bertinetto~\cite{bertinetto2016fully}, Valmadre~\cite{valmadre2017end} and others extended its use to target tracking and obtained satisfactory results. 
\begin{equation}
L_s=\frac{1}{2N}\sum_{n=1}(\|(H^s_a)_n-H^s_v)_n\|^2 ),
\end{equation}
where $n$ is the n-th micro-video; $N$ is the number of samples trained once;   $H^s_a$ and $H^s_v$ are the extracted common domain features of the audio modality and the visual modality, respectively. The similarity representation of the visual modal public domain features and audio modal public domain features is explored by reducing the difference between the visual modal and audio modal in the public domain features at the output of each fully connected layer.

Discrepancy loss: For the variance in distribution, KL scattering algorithm will be used in the experiments to calculate the variance of the same modal distribution since KL scattering is widely used to assess the difference between the predicted and true value distributions of the model output. In the process of reducing the discrepancy loss $LD_KL$, the optimization of the modal private domain network parameters from the perspective of the distribution discrepancy is achieved by the following equation:
\begin{equation}
    L_D=\sum P(H^p_a) log_2 \frac{P(H^p_a)}{Q(H^s_a)}+\sum P(H^p_v) log_2 \frac{P(H^p_v)}{Q(H^s_v)},
\end{equation}
where $P(H^p_a)$ is the probability distribution of the private domain features of audio modality; $Q(H^s_a)$ is the probability distribution of the public domain features of audio modality; $P(H^p_v)$ is the probability distribution of the private domain features of visual modality; $Q(H^p_v)$ is the probability distribution of the public domain features of visual modality. The value of the loss function is larger when the variability of $P(H^p_a)$ and $Q(H^s_a)$ (or $P(H^p_v)$ and $Q(H^p_v)$) is larger, and the value of the loss function is zero when the same.
\section{EXPERIMENTS}
\subsection{Dataset}

The datasets used in this experiment are MovieLens-1M and Amazon.

MovieLens: This dataset is a publicly available movie dataset that contains information about the ratings of different movies by multiple users and the characteristic attributes of users and movies. The dataset can be of various sizes depending on the number of ratings. In this part, we use the movie dataset MovieLens-1M with a sample size of about 1M.

Amazon: This dataset is a publicly available electronic product dataset that contains product reviews and metadata, as well as user ratings and reviews of products.

   \begin{table}
  \centering
  \caption{Rating data format.}
  \label{table_3}
  \setlength{\tabcolsep}{4.0mm}
  \begin{tabular}{|c|c|c|c|}
    \hline
    \textbf{user ID}&\textbf{movie ID}&\textbf{rating}&\textbf{timestamps}\\
    \hline
    \ $1$&$1193$&$5$&$978300760$\\
    \hline
    \ $1$&$661$&$3$&$978302179$\\
    \hline
    \ $\dots$&$\dots$&$\dots$&$\dots$\\
    \hline
    
  \end{tabular}
  \vspace{-2mm}
\end{table}

\subsection{Baselines}

FM~\cite{rendle2010factorization}: Factorization Machine, which simulates first-order feature importance and second-order feature interactions.

DeepFM~\cite{deepfm}: DeepFM is an end-to-end model of a joint decomposer and multilayer sensing machine, which uses deep neural networks and factorization machines to model the interactions of higher-order features and lower-order features, respectively.

ACF~\cite{chen2017attentive}. This is the first framework that is designed to tackle the implicit feedback in multimedia recommendation. It introduces two attention modules to address the item-level and componentlevel implicit feedbacks. To explore the modal-specific user preference and micro-video characteristic, we treat each modality as one component of the micro-video, which is consistent with the idea of standard ACF.

\subsection{Evaluation Metrics}

The loss functions logloss, and AUC are used as the evaluation metrics for this experiment. The log-loss function logloss is calculated as follows:
\begin{equation}
logloss=-\frac{1}{T} \sum_{t=1}^T (y_t log(y_t)+(1-y_t)log(1-y_t)),
\end{equation}

AUC is the area under the ROC curve, which measures the probability that a CFR prediction will score higher for a randomly selected positive sample than for a randomly selected negative sample. The higher the AUC, the better the model performance.

 \begin{figure*}
    \centering
    \subfigure[AUC on Movielens]{
      \includegraphics[width=0.4\textwidth]{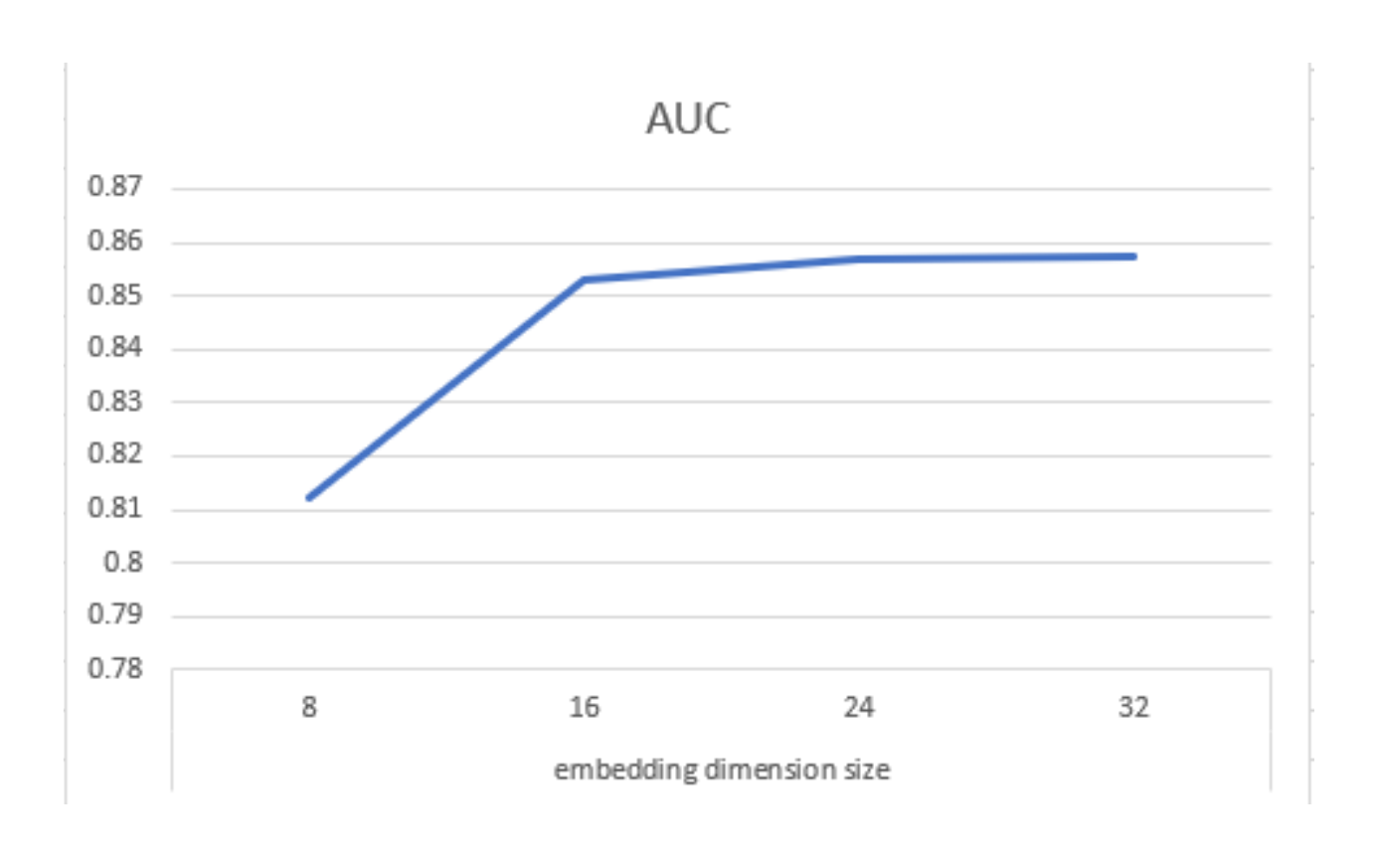}
      \label{fig_visualize_1_1}
    }
    \subfigure[AUC on Amazon]{
      \includegraphics[width=0.4\textwidth]{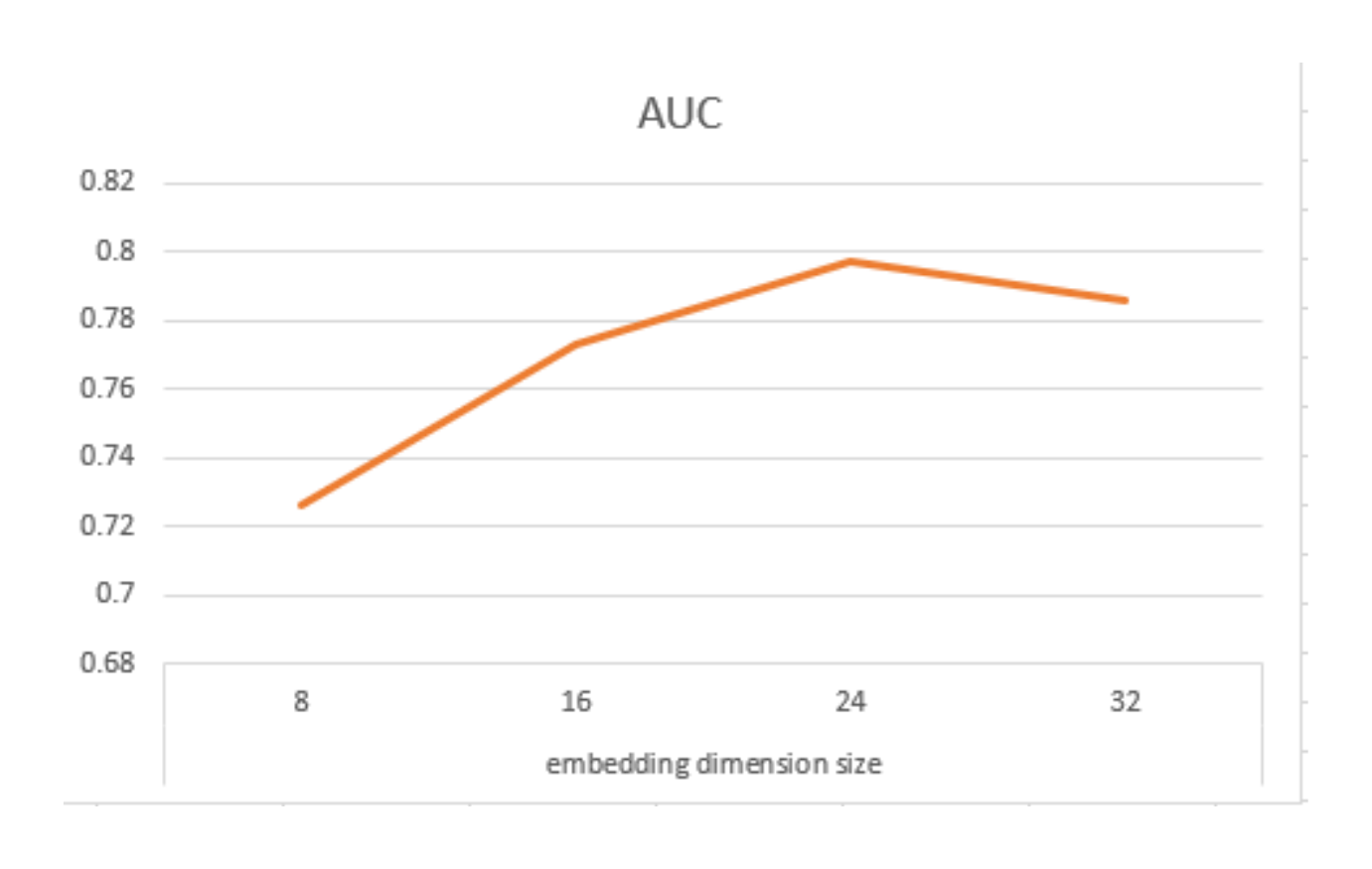}
      \label{fig_visualize_2_3}
    }
    \subfigure[Logloss on Movielens]{
      \includegraphics[width=0.4\textwidth]{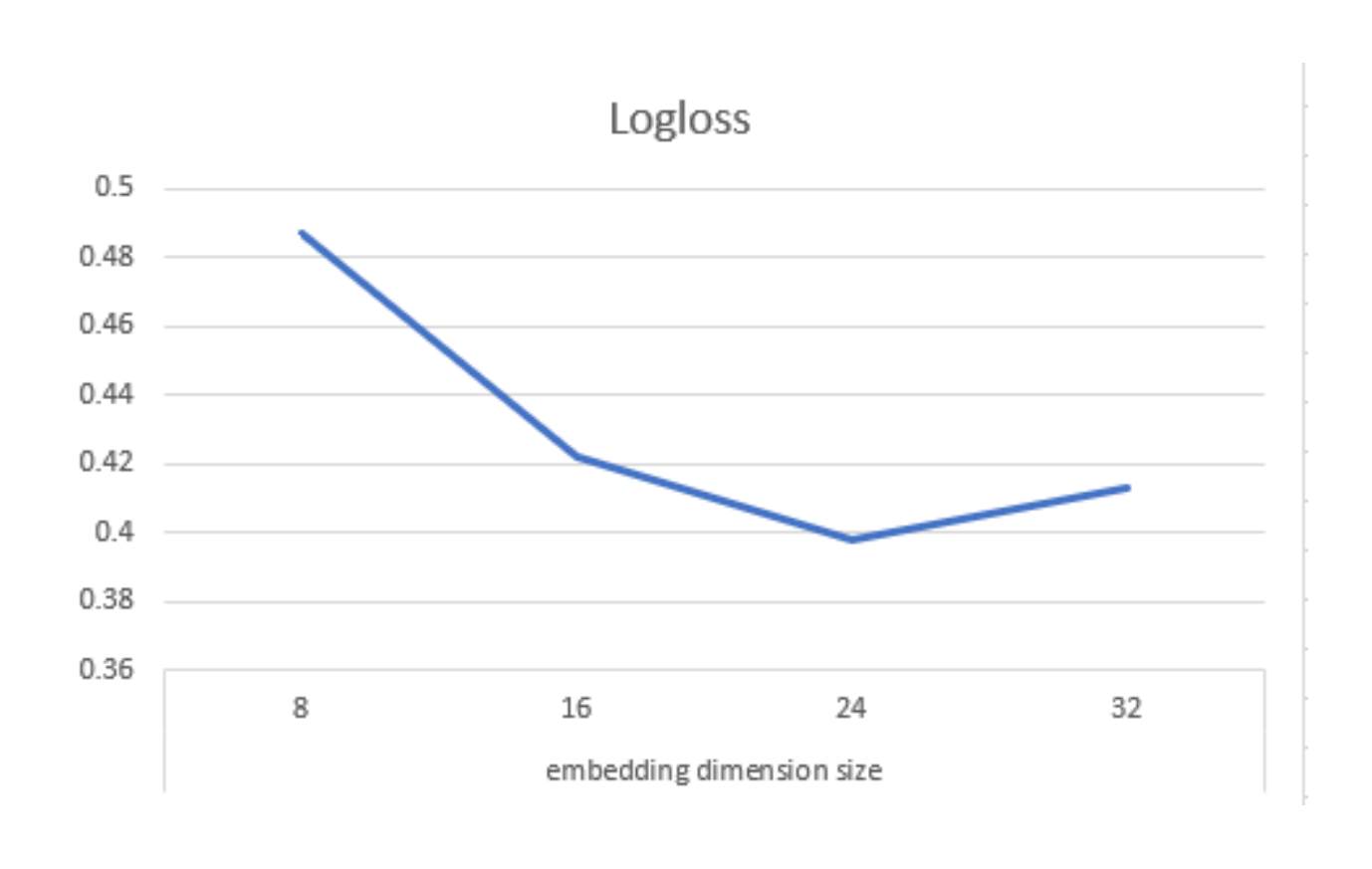}
      \label{fig_visualize_1_1}
    }
    \subfigure[Logloss on Amazon]{
      \includegraphics[width=0.4\textwidth]{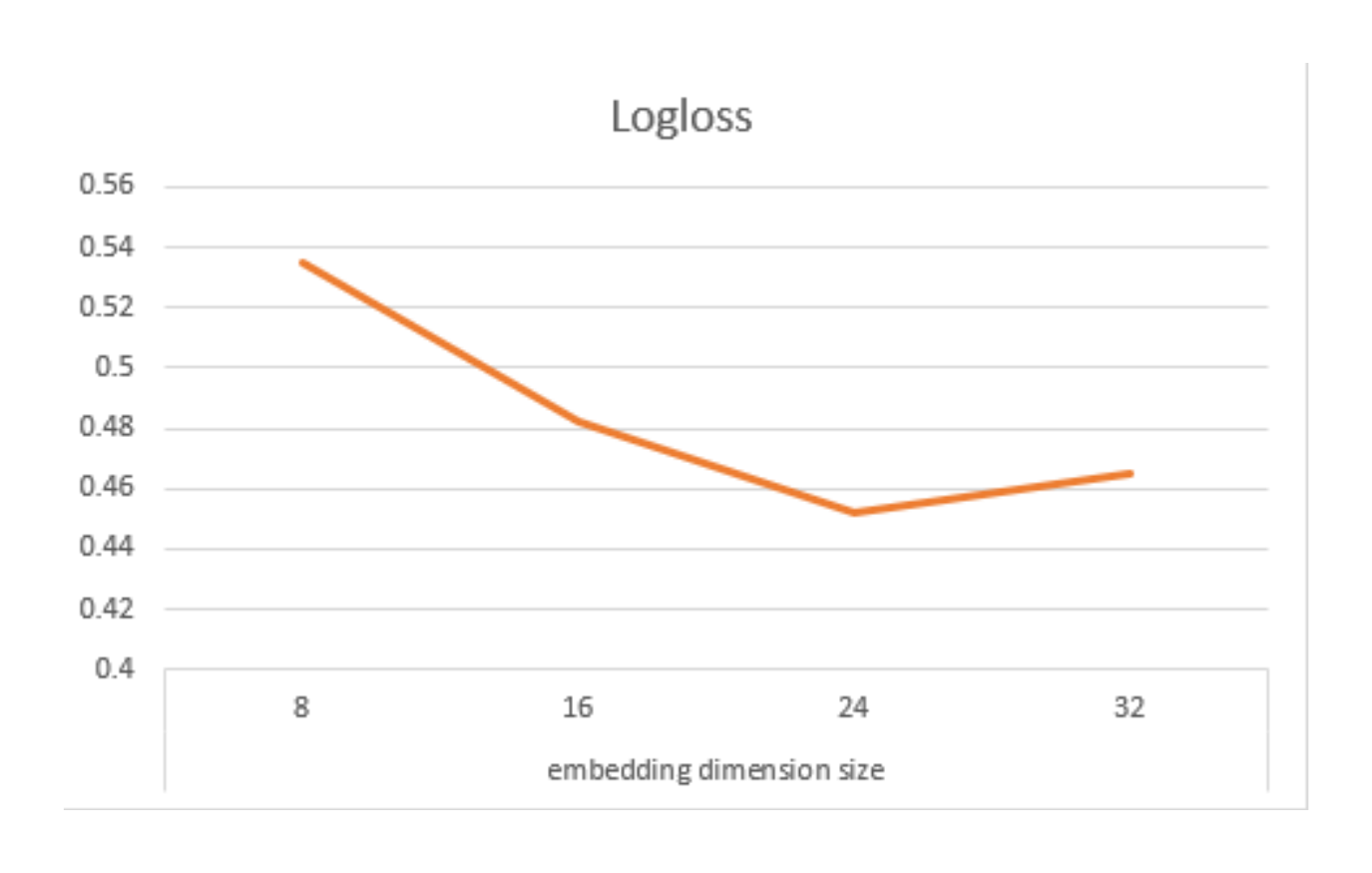}
      \label{fig_visualize_2_3}
    }

    \vspace{-10pt}
    \caption{Effect of embedding dimension size on our model.}
    \label{fig_3}
    \vspace{-5pt}
 \end{figure*}

\subsection{Result Analysis}

The experimental results are shown in Table 2. FM can automatically extract the second-order cross features, but it cannot mine the higher-order feature information, so the performance of the model is lower than that of the DeepFM model. The DeepFM does not get a significant improvement compared to FM, which is due to the fact that the training parameters of the model increase after the introduction of the Deep part, however, the data set is smaller, so the Deep part of the model is limited. When the model uses only the basic interaction data, the model proposed in this paper performs better than both the FM model and the DeepFM model. This also demonstrates the strong expressive power of the two-layer feature representation model proposed in this paper. The reason for this is that our model considers both the internal correlation between features and the importance of feature cross-correlation. Therefore, compared with the FM model, our model has a stronger feature representation and better generalization of the model. By adding the MLP structure to our model, higher-order feature interaction information is mined. The model proposed in this paper adds multimodal video content features as well as social features to the basic interaction features. The results of the comparison experiments on the dataset show that our model with the addition of multimodal content and social information is more beneficial than the recommendation effect using basic interaction data alone in both LogLoss and AUC metrics. The main reason is that multimodal video content and social information provide richer and more comprehensive descriptions to video recommendations, and at the same time, the use of user-guided attention networks to fuse multimodal video content enables the model to learn personalized video content features that satisfy user preferences, thus improving the accuracy of the model.
\begin{itemize}
    \item The fusion method by directly splicing video content features of each modality including visual features, audio features, and text features results in a $0.86\%$ reduction in the LogLoss metric and a $1.16\%$ improvement in the AUC metric for the reason that the inclusion of multimodal video content can provide more information and thus improve the model expression capability. 
    \item (2) The attention-based multimodal feature fusion method leads to a $0.88\%$ reduction in the LogLoss metric and a $1.2\%$ improvement in the AUC metric for the experimental results.
    \item By adding the extracted crossover features, it makes the model reduce $1.4\%$ on the indicator LogLoss and improve $2.01\%$ on the indicator AUC. The experimental results show that the crossover features help to predict users' preferences and have a significant effect on improving the prediction ability of the model, which also verifies the importance of crossover features in recommendation systems.

\end{itemize}

   \begin{table}
  \centering
  \caption{Model performance comparison.}
  \label{table_3}
  \setlength{\tabcolsep}{8.0mm}
  \begin{tabular}{|c|c|c|}
    \hline
    \ &\textbf{AUC}&\textbf{Logloss}\\
    \hline
    \textbf{FM} & 0.6548 &0.5766\\
    \hline
    \textbf{DeepFM} & 0.6632 &0.5712\\
    \hline
    \textbf{ACF} & 0.6532 &0.5697\\
    \hline
    \textbf{Ours} & 0.6976 &0.5583\\
    \hline
    
  \end{tabular}
  \vspace{-2mm}
\end{table}
\section{Conclusion and Future Work}
In the application scenario of video recommendation, users pay different attention to each modality such as text, audio and visual in the video. Therefore, in this paper, we propose an attention-based multimodal feature fusion method to assign different weights to each modality of the video through an attention mechanism. In order to ensure that the learned weights of each modality can reflect the user's preferences, the method incorporates user features into the attention network to achieve personalized fusion of the features of each modality of the video, thus providing more personalized video recommendations. To address the main micro-video classification tasks, by establishing similarity loss and difference loss, we explore the similarity between different modalities in micro-videos and the difference of the same modality, to obtain the private domain features and public domain features of different modalities of the video, and fuse them as the global features, and the classification loss is used to guide the classification of micro-videos.Extensive experiments conducted on a real-world dataset have well-verified that our model significantly outperforms several state-of-the-art baselines.

\bibliography{BIB/IEEEabrv, reference}

\end{document}